\documentclass[conference]{IEEEtran}
\IEEEoverridecommandlockouts
\usepackage{cite}
\usepackage{amsmath,amssymb,amsfonts}
\usepackage{algorithmic}
\usepackage{graphicx}
\usepackage{textcomp}
\usepackage{subcaption}
\usepackage[dvipsnames]{xcolor}
\usepackage{float}
\usepackage{balance}
\usepackage{listings}
    \definecolor{vgreen}{RGB}{104,180,104}
    \definecolor{vblue}{RGB}{49,49,255}
    \definecolor{vorange}{RGB}{255,143,102}
    \lstdefinestyle{verilog-style}
    {
        language=Verilog,
        basicstyle=\small,
        keywordstyle=\color{vblue},
        identifierstyle=\color{black},
        commentstyle=\color{vgreen},
        numbers=left,
        numberstyle={\tiny \color{black}},
        numbersep=10pt,
        tabsize=8
    }
\def\BibTeX{{\rm B\kern-.05em{\sc i\kern-.025em b}\kern-.08em
    T\kern-.1667em\lower.7ex\hbox{E}\kern-.125emX}}
    
    \newcommand{\rulesep}{\unskip\ \vrule\ }

\usepackage{xcolor}
\usepackage{circledsteps}
\usepackage{xspace}
\usepackage{tcolorbox}
\tcbuselibrary{listings}
\newcommand\myCircled[2][]{\ifmmode
\Circled[fill color=black,inner color=white,#1]{\mathsf{#2}}
\else
\Circled[fill color=black,inner color=white,#1]{\sffamily#2}
\fi
}    
    
\usepackage{background}
\usepackage[usestackEOL]{stackengine}
\setstackgap{L}{\normalbaselineskip}
\SetBgContents{\color{blue}{\tiny \Longstack{PREPRINT - accepted at  21st IEEE Interregional NEWCAS Conference.}}}
\SetBgPosition{4.5cm,1cm}
\SetBgOpacity{1.0}
\SetBgAngle{0}
\SetBgScale{1.8}

\begin{document}
\bstctlcite{IEEEexample:BSTcontrol}
\title{Automated Information Flow Analysis for Integrated Computing-in-Memory Modules\vspace{-0.25cm}}

\author{Lennart M. Reimann, Felix Staudigl and Rainer Leupers\\
RWTH Aachen University\\
\{lennart.reimann, staudigl, leupers\}@ice.rwth-aachen.de\\
\vspace{-0.9cm}
}

\maketitle

\begin{abstract}
Novel non-volatile memory (NVM) technologies offer high-speed and high-density data storage. In addition, they overcome the von Neumann bottleneck by enabling computing-in-memory (CIM). Various computer architectures have been proposed to integrate CIM blocks in their design, forming a mixed-signal system to combine the computational benefits of CIM with the robustness of conventional CMOS. Novel electronic design automation (EDA) tools are necessary to design and manufacture these so-called neuromorphic systems. Furthermore, EDA tools must consider the impact of security vulnerabilities, as hardware security attacks have increased in recent years. Existing information flow analysis (IFA) frameworks offer an automated tool-suite to uphold the confidentiality property for sensitive data during the design of hardware.
However, currently available mixed-signal EDA tools are not capable of analyzing the information flow of neuromorphic systems. To illustrate the shortcomings, we develop information flow protocols for NVMs that can be easily integrated in the already existing tool-suites. We show the limitation of the state-of-the-art by analyzing the flow from sensitive signals through multiple memristive crossbar structures to potential untrusted components and outputs. Finally, we provide a thorough discussion of the merits and flaws of the mixed-signal IFA frameworks on neuromorphic systems.
\end{abstract}

\begin{IEEEkeywords}
information flow analysis, neuromorphic computing, confidentiality 
\end{IEEEkeywords}

\section{Introduction}
Non-volatile memory (NVM) technologies, such as spin-torque transfer memory (STT-RAM/MRAM), phase-change random access memory (PCRAM), or redox-based random access memory (ReRAM), are promising candidates to substitute traditional RAM. NVMs offer dense storage with low leakage power, and enable computing-in-memory.

\begin{figure}[t]
    \centering
    \includegraphics[width=\columnwidth]{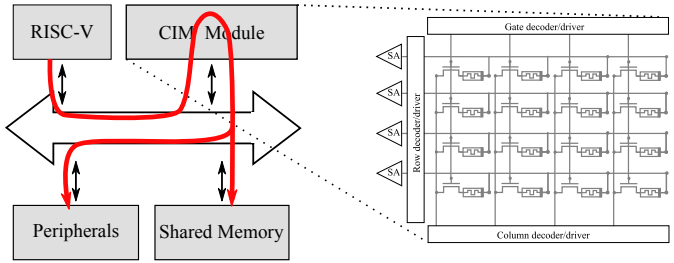}
    \caption{Exemplary data leakage paths in a SoC using a CIM module (1T1R crossbar). Sensitive data flows from the trusted source (RISC-V) to untrusted peripherals or shared memories. \label{fig:motivation}\vspace{-0.9cm}}
    
\end{figure}

Combining conventional CMOS with NVMs results in complex high-performance designs referred to as neuromorphic systems. In modern design processes, electronic design automation (EDA) tools are used to assist the designer in the intricate implementation. However, the EDA tools need to be equipped to facilitate mixed-signal designs incorporating NVM-based accelerators. Furthermore, novel technologies introduce new security vulnerabilities (see Fig.~\ref{fig:motivation})~\cite{caches, data_leakage, memristor_attacks}. These new vulnerabilities are particularly worrying, because neuromorphic systems are a promising candidate for future applications, such as autonomous driving. Consequently, EDA tools need to be adapted to enforce security properties for mixed-signal designs, enabling a security-aware design flow in both the digital and analog domain. Availability, confidentiality, and integrity are the three cornerstones of hardware security that must be considered during the design process. While most research focuses on the integrity property~\cite{neurohammer, xfault}, we concentrate in our work on the confidentiality property. Information flow analysis (IFA) is the state-of-the-art technique to enforce the confidentiality and integrity property in a design. IFA can track the flow of information from sensitive sources, such as encryption keys, to untrusted components, such as outputs or third-party intellectual property (IP). The analysis can be conducted statically to ensure confidentiality of a signal for every possible input combination. Although some work has been published supporting mixed-signal designs, no work has been released to the best of our knowledge that discusses the information flow for NVMs. Consequently, our work extends currently available IFA tools for neuromorphic mixed-signal systems and discusses its merits and shortcomings.

The major contributions of this paper are:
(1) Development of information flow policies for NVM components. (2) An introduction of a crossbar driver masking scheme to \textit{forbid} sneak paths in hardware. (3) A demonstration of the limitations of current mixed-signal IFA tools for neuromorphic systems.

\section{Background}
\subsection{Information Flow Analysis}
Information flow analysis represents the state-of-the-art technique to enforce the integrity and confidentiality property in a hardware design. The security analysis requires the hardware to be divided and labeled in different security classes. For instance, third-party IP, shared resources, or the output ports of the design are labeled untrustworthy. IFA determines whether information from high-security parts affects lower-security areas. The analysis relies on the non-interference property, aiming to prove that a change in sensitive values does not lead to an observable change in the untrustworthy components. The sensitive signals do not interfere with insecure components. Enforcing the properties during every step of the design process, avoiding security vulnerabilities that threaten sensitive data, such as encryption keys or user data. 

\subsection{VeriCoq-IFT}
The majority of IFA frameworks are designed to handle digital hardware. In contrast, the VeriCoq-IFT framework~\cite{vericoq_ift} introduces the capabilities to process mixed-signal designs when analyzing the information flow~\cite{vericoq_mixed_signals}. 
The analysis indicates whether sensitive information is leaked to the design's output signals. First, all output ports of the design must be labeled untrustworthy. Second, the user marks the sensitive signal in the design description and assigns it a sensitivity score. The conservative approach of VeriCoq-IFT propagates the sensitivity score of a signal at every signal assignment in Verilog. A variable receives the highest sensitivity score of all its inputs. Furthermore, operations can be labeled a \textit{sensitivity reducer}, so that every time the sensitive signal passes the designated operation, the sensitivity score is reduced. If the signal reaches an output before it reaches a sensitivity score of zero, a data leakage is detected. The score system can be used to enforce that, e.g., the plaintext passes an AES round at least 12 times before it reaches the design's output as the ciphertext. Nevertheless, information flow rules for memristors have not been introduced yet.

\subsection{Non-volatile memories (NVMs)}
NVMs represent a novel memory technology that takes advantage of the memristors.  A memristor is next to a capacitor, a resistor, and an inductor, the fourth fundamental electrical component and stores information in the form of resistance. The resistance which can be set or reset using voltage pulses represents different states, which are called low resistive state (LRS) and high resistive state (HRS). To achieve high densities, memristors are organized in crossbar structures consisting of horizontal word and vertical bit lines, with a memristive cell at each cross point. These so-called passive crossbars suffer from sneak-path currents based on parasitic effects between the memristive cells limiting their reliability and retention. Consequently, more advanced cells have been proposed incorporating an active component, i.e., transistor, to connect/isolate the memristor from the remaining crossbar.
However, as these novel devices have not been discussed in the VeriCoq-IFT framework yet~\cite{vericoq_mixed_signals}, information flow policies need to be developed and implemented to enable the analysis of information flows of neuromorphic systems.
\begin{figure}[t]
    \centering
    \begin{subfigure}[b]{0.45\columnwidth}
         \centering
    \includegraphics[origin=c, width=0.8\columnwidth]{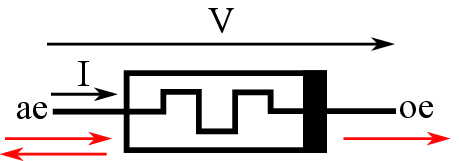}
    \caption{Set/Read}
    \label{fig:flow_policies_set}
    \end{subfigure}
    \rulesep
    \begin{subfigure}[b]{0.45\columnwidth}
         \centering
    \includegraphics[origin=c, width=0.8\columnwidth]{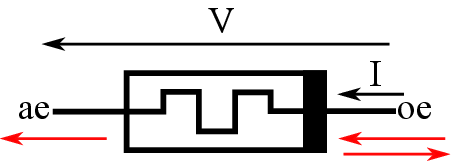}
    \caption{Reset}
    \label{fig:flow_policies_reset}
    \end{subfigure}
    \caption{Information flow (red arrows) in a memristor for different operational modes. \label{fig:flow_policies}\vspace{-0.4cm}}
   
\end{figure}

\section{Threat Model}
The developed framework aims to identify undesired information leakages in neuromorphic mixed-signal designs. The static analysis is conducted on register transfer level for the digital components, and transistor-level for the analog parts.
We assume the attacker has access to the complete hardware description and intends to leak information via a direct flow of information at the primary outputs, no matter whether those outputs lie in the analog or digital domain. Side-channels are not considered. We assume the hardware vulnerabilities are already present at the design stage. It is not considered whether the observations of the primary outputs are obtained via remote access or physical access.

\section{Related Work}
Khan et al. elaborate on possible attacks on information leaks on emerging non-volatile memories by using side channels caused by supply noise when writing and reading sensitive data \cite{side_channels}. 
Furthermore, current research has shown that data-dependent write latencies can be exploited as a side-channel to leak sensitive information. By observing the time to access a memristor, information about the current content can be derived. The analysis regarding this vulnerability has also been conducted manually~\cite{data_leakage}.
In addition to side-channels through the supply noise and the write latency, the supply current can be observed to gather information about sensitive signals\cite{caches}.

Although multiple vulnerabilities in NVMs have been identified, no work has been presented to automate the identification of such vulnerabilities. Automated security-aware EDA tools are required to assist a hardware designer, inexperienced in hardware security, in identifying security vulnerabilities while maintaining a competitive design process.

\section{Framework}
Therefore, we develop information flow rules for NVMs and integrate them in known IFA frameworks. Fig.~\ref{fig:flow_policies} illustrates the direction of voltage and current for the three operational modes of a memristor: \textit{set}, \textit{read} and \textit{reset}. For \textit{set} and \textit{read} the memristor is accessed from the terminal $ae$, so the current and voltage directions aim at the other terminal, called $oe$ in this work. Fundamentally, the memristor requires a fourth mode to initialize the device after manufacturing. However, we do not consider this mode in our work because of its limited attack surface compared to the remaining other operational modes.

As stated in ~\cite{vericoq_mixed_signals}, for analog components, information is carried by voltage \textit{and} current. Therefore, when setting the voltage at $ae$, the information can be read at both terminals of the memristor via the current. Due to the bidirectional behavior of a memristor~\cite{reram_switching}, the information flow behaves bidirectionally, as illustrated with the red arrows in Fig.~\ref{fig:flow_policies} for the three access modes. 

\begin{figure}[t!]
    \centering
    \includegraphics[width=\columnwidth]{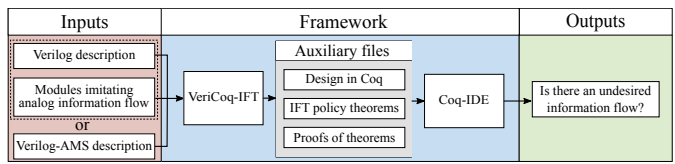}
    \caption{Functionality of VeriCoq-IFT~\cite{vericoq_ift} and the CoqIDE~\cite{coq_ide} in this work.\vspace{-0.6cm}
    \label{fig:vericoq_ift}}
\end{figure}
\begin{figure}
\begin{lstlisting}[style={verilog-style}]
// Mimicking the information flow
// in memristors
module memristor (ae, oe);
    inout ae, oe;
    
    assign ae = ae | oe;
    assign oe = ae | oe;
endmodule
\end{lstlisting}
\caption{High-level definition of a Verilog memristor module modeling the information flow.\label{fig:verilog_memristor}\vspace{-0.6cm}}

\end{figure}
\begin{figure*}[ht!]
    \centering
    \includegraphics[width=\textwidth]{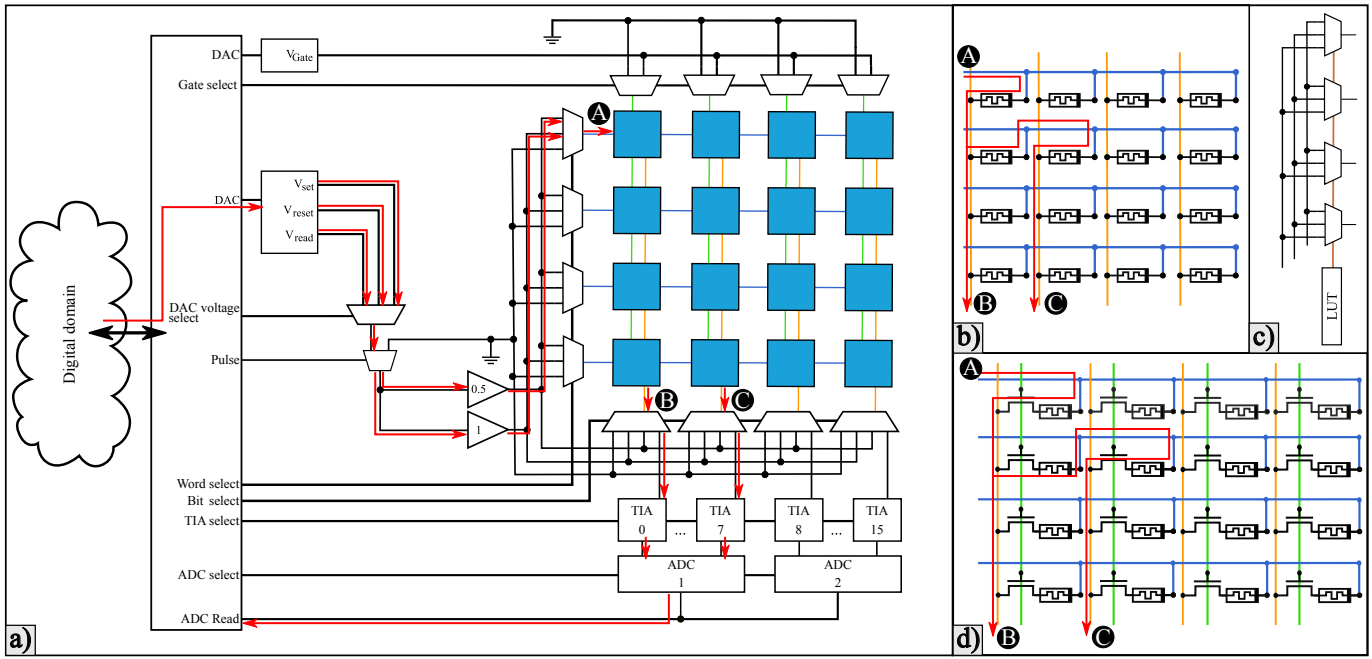}
    \caption{The demonstration setup: (a) The computing-in-memory module and its interface to the digital domain with a generic crossbar (blue), (b) a 1R crossbar that can be integrated in the CIM on the left for the generic crossbar, (c) a LUT for the drivers that forbids certain voltage combinations, and (d) a 1T1R crossbar that can be integrated in the CIM module on the left.\label{fig:crossbar_leakages} The red lines indicate exemplary leakages for a sensitive signal in the digital domain, identified by our VeriCoq-IFT setup.\vspace{-0.4cm}}
\end{figure*}

We integrated the derived policies in VeriCoq-IFT and combined the framework directly with the CoqIDE~\cite{coq_ide} to provide an automated IFA framework. The tool flow of the combined VeriCoq-IFT and CoqIDE framework used in this work for the evaluation is depicted in Fig.~\ref{fig:vericoq_ift}. Although VeriCoq-IFT was introduced for third-party IP as proof-carrying hardware IP, it is solely used for the IFA in this work. VeriCoq-IFT has two operational modes: 1) It processes the Verilog-A/MS description of the complete design \textit{or} 2) The designer provides a Verilog description of the digital domain, combined with Verilog modules of the analog modules that mimic the information flow of the device. The latter is required for the early design stages, when no Verilog-A/MS of the memristor is yet available. Then, Verilog modules mimicking the information flow of a memristor need to be introduced. The model does not depict the actual behavior of a NVM device, but models the information flow. Fig.~\ref{fig:verilog_memristor} depicts the Verilog module mimicking the information flow of a memristor. Both terminals $ae$ and $oe$ are labeled 'inout'-ports. Additionally, each of the two ports depends on both terminals (line 6 \& 7). The type of operation performed on the right side of the Verilog assignment is irrelevant, as only a flow of information needs to be modeled, not the functionality. This allows the framework to handle analog memristors in the digital domain. 

Secondly, modern EDA tools allow the export of mixed-signal design into the language Verilog-A/MS. The analog and digital behavior of the components is embedded into a single description. A small number of Verilog-A/MS devices are available online~\cite{ams_models}. VeriCoq-IFT processes the design to generate a design description in the language Coq, and theorems and proofs of the information flow rules. These rules are generated for all signals labeled sensitive in the design description. In this work, \textit{all sensitivity labels are set to 1 and no sensitivity reducers are instantiated}, enforcing the non-interference property~\cite{non_interference_property}. Therefore, every output port that can be influenced by the sensitive signal is labeled a leakage point. The three auxiliary files are forwarded to the theorem prover in the CoqIDE. If the theorems and proofs pass for the hardware description in Coq, no leakage is detected, otherwise VeriCoq-IFT identifies undesired flows of information.

\section{Demonstration}
In this work, we present the functionality and limitations of the presented framework using three individual integrated NVM-based mixed-signal designs. We assume the NVM block is integrated on a system-on-chip (SoC) accelerator. We focus our evaluation on the analog domain to illustrate the capabilities and shortcomings of the implemented framework. Fig.~\ref{fig:crossbar_leakages} illustrates the analog domain of the SoC and marks one identified leakage path in red. The surrounding circuitry enables in this example design the orchestration of both passive and active crossbar arrays.
Following, we conduct three experiments to highlight the obstacles of IFA to neuromorphic systems and ultimately the shortcomings of the IFA framework. The internals of both crossbar structures are shown in Fig.~\ref{fig:crossbar_leakages} (b) and (d). The two crossbar structures can each be integrated into the circuitry (Fig.~\ref{fig:crossbar_leakages}(a)) by replacing the blue abstract crossbar. Furthermore, we propose a masking mechanism that enforces the intended usage of the NVM module (see Fig.~\ref{fig:crossbar_leakages} (c)), which would replace the drivers shown in Fig.~\ref{fig:crossbar_leakages} (a).

\subsection{1R Crossbar Accelerator}
The memory cell of a passive crossbar consists of a single memristor. Hence, the crossbar itself acts like a network of resistances, allowing a bidirectional information flow. The digital domain limits the information flow based on the implemented output signals, i.e., to compute a vector-matrix multiplication, or to communicate the result by an input signal. Fig.~\ref{fig:crossbar_leakages} (b) exemplifies in red one possible information leakage path additionally to the intended flow of information. In addition to the intended flow of information from \myCircled{A} to \myCircled{B}, the undesired sneak paths are identified too (see \myCircled{A} to \myCircled{C})~\cite{sneak_paths}. Overall, the information flow in passive crossbars is considered complex, since there is no clear direction and the information can be transferred alternating between the analog and digital domain.

\subsection{1T1R Crossbar Accelerator}
While passive crossbar introduces sneak path currents limiting their usability, active crossbar aims to solve this by extending the NVM cell by an active component~\cite{sneak_paths}. Fig.~\ref{fig:crossbar_leakages} (d) illustrates an active crossbar using a transistor as a selector component to separate unselected cells from the crossbar. However, our experiments show that the information flow, determined by VeriCoq-IFT, of an active crossbar matches the flow of a passive crossbar. \textit{VeriCoq-IFT performs a static analysis of the design which does not take into account the ``intended" usage of the selector transistors. Consequently, the framework classifies the NVM module as leaky.}

\subsection{Active Crossbar Accelerator with Access Mask}
To secure the information flow through crossbar arrays, we propose a hard-wired \textit{access mask} enforcing the intended usage of the selector transistors. We implement this \textit{access mask} within the driver circuitry of the NVM block. The mask allows only a fixed set of driver voltages to be applied to the crossbar, as shown in Table~\ref{tab:masks} and is implemented with a lookup table (see Fig.~\ref{fig:crossbar_leakages} (c)). The lookup table forbids operational modes that can simultaneously write to multiple rows, thus omitting sneak paths. For instance, if the memristor in row $m$ and column $n$ (green) is accessed, all other word and bit lines are set to $GND$ (orange), blocking potential sneak paths. \textit{The results of the VeriCoq-IFT analysis are illustrated with the red leakage paths (Fig.~\ref{fig:crossbar_leakages} (d)), which are the same as for the previous two hardware designs.}

\begin{table}[b!]
  \centering
  \vspace{-0.4cm}
    \caption{Rules for the allowed access masks\label{tab:masks} to set or reset the crossbar at ($k$,$l$) when using a ($m$,$n$)-crossbar.}
    \begin{tabular}{r|r}
       \multicolumn{1}{c|}{\textbf{SET}}  & \multicolumn{1}{c}{\textbf{RESET}} \\\hline
        \textcolor{OliveGreen}{$V_{WL,k} = V_{SET}$} &  \textcolor{OliveGreen}{$V_{WL,k} = V_{RES}$}\\
        \textcolor{OliveGreen}{$V_{SL,l} = V_{GAT}$} &  \textcolor{OliveGreen}{$V_{SL,l} = V_{GAT}$}\\
        \textcolor{OliveGreen}{$V_{BL,l} = GND$} &  \textcolor{OliveGreen}{$V_{BL,l} = GND$}\\
        \textcolor{BrickRed}{$V_{WL,1:k-1} = GND$} &  \textcolor{BrickRed}{$V_{WL,1:k-1} = GND$}\\
        \textcolor{BrickRed}{$V_{SL,1:l-1} = GND$} &  \textcolor{BrickRed}{$V_{SL,1:l-1} = GND$}\\
        \textcolor{BrickRed}{$V_{BL,1:l-1} = GND$} &  \textcolor{BrickRed}{$V_{BL,1:l-1} = GND$}\\
        \textcolor{BrickRed}{$V_{WL,k+1:m} = GND$} &  \textcolor{BrickRed}{$V_{WL,k+1:m} = GND$}\\
        \textcolor{BrickRed}{$V_{SL,l+1:n} = GND$} &  \textcolor{BrickRed}{$V_{SL,l+1:n} = GND$}\\
        \textcolor{BrickRed}{$V_{BL,l+1:n} = GND$} &  \textcolor{BrickRed}{$V_{BL,l+1:n} = GND$}\\
    \end{tabular}

\end{table}
\section{Discussion}
The VeriCoq-IFT analysis of the 1R crossbar accurately depicts the behavior of the sneak paths, so that the framework identifies possible data leakages. On the contrary, active crossbars solve the influence of sneak path currents by incorporating active components, such as transistors. As a static analysis cannot depict the difference between active and passive crossbars, the information flow did not change. Although an access mask in the control circuitry must lead to a change in the identified information flow, VeriCoq-IFT's conservative analysis is not capable of differentiating between the three designs' flows, leading to false positive identifications. Specifically, the SoC designs with a CIM module, which allows continuous information flow between the analog and digital domain, require a less conservative approach to reduce the high number of false positives. Less conservative approaches are already available for the digital domain~\cite{sec_verilog}, but are yet missing for the analog domain. The accurate frameworks consider inter-signal dependencies, the actual functionality of an operation, and the accurate value of a signal, which is not done by VeriCoq-IFT. Thus, a framework needs to be developed that considers the mentioned features in the analog by processing the information in the Verilog-A/MS description.

\section{Conclusion}
This work presented an evaluation of the state-of-the-art information flow analysis framework VeriCoq-IFT for integrated CIM modules. As demonstrated, derived Verilog models could be implemented to enable an early stage IFA for trending memristor crossbars, a crucial building block for neuromorphic systems. The functionality of the mixed-signal IFA was demonstrated using three system designs. However, the conservative nature of the current analog information flow theorems leads to many false positives, which make a practical static analysis of the confidentiality property in CIM modules infeasible. In future work, the framework could be extended to allow a less conservative analysis of the information flow~\cite{sec_verilog} or even a quantification of the information flow, so that negligible flows can be ignored~\cite{qflow, qflow2}.

\section*{Acknowledgment}
The work was funded by the German Federal
Ministry of Education and Research (BMBF) within the project
NEUROTEC II under contract no. 16ME0399. The VeriCoq framework was provided by the TRELA laboratory at UT Dallas.

\clearpage
\balance
\bibliographystyle{IEEEtran}
\bibliography{bibtexentry}

\end{document}